\documentstyle[psfig]{caosp}
\begin{document}
%
%
%%%%%%%%%%%%%%%%%%%%%%%%%%%%%%%%%%%%%%%%%%%%%%%%%%%%%%%%%%%%%%%%%%%%%%%%%%%%%%
%\pubyear{1998}
%\volume{27}
%\firstpage{368}
%%%%%%%%%%%%%%%%%%%%%%%%%%%%%%%%%%%%%%%%%%%%%%%%%%%%%%%%%%%%%%%%%%%%%%%%%%%%%%
%%%%%%%%%%%%%%%%%%%%%%%%%%%%%%%%%%%%%%%%%%%%%%%%%%%%%%%%%%%%%%%%%%%%%%%%%%%%%%
%\htitle{Ultraviolet Spectral Atlas of $\alpha$$^2$ CVn}
\hauthor{Glenn M. Wahlgren}
%%%%%%%%%%%%%%%%%%%%%%%%%%%%%%%%%%%%%%%%%%%%%%%%%%%%%%%%%%%%%%%%%%%%%%%%%%%%%%
\title{On an ultraviolet spectral atlas of $\alpha$$^2$ CVn}
%%%%%%%%%%%%%%%%%%%%%%%%%%%%%%%%%%%%%%%%%%%%%%%%%%%%%%%%%%%%%%%%%%%%%%%%%%%%%%
\author{Glenn M. Wahlgren}
%%%%%%%%%%%%%%%%%%%%%%%%%%%%%%%%%%%%%%%%%%%%%%%%%%%%%%%%%%%%%%%%%%%%%%%%%%%%%%
\institute{Atomic Spectroscopy Group\\Department of Physics, University of 
Lund\\S\"{o}lvegatan 14, S-223 62, Lund, Sweden}
%%%%%%%%%%%%%%%%%%%%%%%%%%%%%%%%%%%%%%%%%%%%%%%%%%%%%%%%%%%%%%%%%%%%%%%%%%%%%%
%                        D A T E                                             %
% Date inserted here will be the date of your paper receiving                %
% or you can use \today command instead explicit date in brackets            %
%%%%%%%%%%%%%%%%%%%%%%%%%%%%%%%%%%%%%%%%%%%%%%%%%%%%%%%%%%%%%%%%%%%%%%%%%%%%%%
\date{\today}
\maketitle
%%%%%%%%%%%%%%%%%%%%%%%%%%%%%%%%%%%%%%%%%%%%%%%%%%%%%%%%%%%%%%%%%%%%%%%%%%%%%%
%                        A B S T R A C T,  K E Y W O R D S                   %
% Here is shown how to write the abstract                                    %
% Write your keywords using command \keywords  the thesaurus from Astron.    %
% Astrophys. Abstracts.                                                      %
%%%%%%%%%%%%%%%%%%%%%%%%%%%%%%%%%%%%%%%%%%%%%%%%%%%%%%%%%%%%%%%%%%%%%%%%%%%%%%
\begin{abstract}
We describe the ultraviolet spectra of the Ap star $\alpha^2$ CVn that have 
been acquired with the Hubble Space Telescope and the nature of the optical
region analyses for its rotational velocity and elemental abundances that have
been undertaken in support of the ultraviolet data.
\end{abstract}
%%%%%%%%%%%%%%%%%%%%%%%%%%%%%%%%%%%%%%%%%%%%%%%%%%%%%%%%%%%%%%%%%%%%%%%%%%%%%%
%                       S E C T I O N I N G                                  %
% A section starts with the command \section as shown below, with the title  %
% in Initial Capitals and lowercase only. Do not number the sections - let   %
% LaTeX do that for you - and do not end them by ".".                        %
%%%%%%%%%%%%%%%%%%%%%%%%%%%%%%%%%%%%%%%%%%%%%%%%%%%%%%%%%%%%%%%%%%%%%%%%%%%%%%
\section{Introduction}
%%%%%%%%%%%%%%%%%%%%%%%%%%%%%%%%%%%%%%%%%%%%%%%%%%%%%%%%%%%%%%%%%%%%%%%%%%%%%%
%                       L A B E L                                            %
% Label command is very convenient for you when referring to secctions,      %
% subsections,..., tables, figures as well as equations in your article (see %
% commands \ref and \pageref). In case figure and table environments         %
% the \label command should always be put after the \caption command to      %
% preserve a proper numbering. When using the \label command you must compile%
% the file twice to get proper cross-references.                             %
%%%%%%%%%%%%%%%%%%%%%%%%%%%%%%%%%%%%%%%%%%%%%%%%%%%%%%%%%%%%%%%%%%%%%%%%%%%%%%
\label{intr}
The spectrum of the prototypical magnetic star $\alpha$$^2$ CVn (= HR 4915,
A0p Si, Hg, Eu, Cr) has been recognized as peculiar for one hundred 
years (Maury 1897).  As a result of its brightness it has been 
an important target for several lines of inquiry into warm
chemically peculiar stars. Yet even though
$\alpha^2$ CVn has been a centerpiece for Ap star study, there is still much
to learn from its spectrum, particularly at ultraviolet (UV) wavelengths. We
have undertaken to study its UV spectrum in order to more fully identify
those ions that are present and better quantify the rich line opacity that
leads to flux redistribution.  
The UV atlas consists of 35 wavelength settings of the Goddard
High Resolution Spectrograph (GHRS) onboard the Hubble Space Telescope
(HST). Three first-order 
gratings, characterized by spectral resolving powers between  
R = $\lambda$/$\Delta\lambda$ = 20000 and 35000, were used to collect
high quality (S/N $>$ 100) data over the wavelength interval 1510 to 2640 \AA.
The data were collected on three separate dates, each corresponding
to magnetic phase 0.0 +/- 0.1, the phase of maximum UV opacity and 
Eu\,{\sc ii} line strength at optical wavelengths.

However, the nature of the UV spectrum can not be studied independently 
of the optical region, and it was deemed necessary to acquire optical region 
spectra to address issues such as rotational velocity and elemental abundances. 
Optical region spectra were obtained during several observing campaigns with
the 2.6-m Nordic Optical Telescope (NOT). The SOFIN echelle spectrograph
was used to obtain complete optical coverage at a resolving power of R = 25000
at a single phase, along with limited wavelength coverage observations for many 
phases at a resolving power of R = 80000.

\section{Optical region analyses}
\subsection{Rotational velocity}

The projected equatorial rotational velocity, $v\sin i$, was determined by 
synthetic
spectrum fitting of unblended Fe\,{\sc ii} lines found in the red spectral
region. The use of Fe\,{\sc ii} lines essentially minimizes line broadening
effects arising from hyperfine structure (hfs) and isotope shifts (IS). No
account was made of the possible Zeeman broadening, so the result may be
considered an upper limit. Line breadths were first compared for
the observed rotational phases for each of several ions
(Fe\,{\sc ii}, Si\,{\sc ii}, Eu\,{\sc ii}, Cr\,{\sc ii}) in order to identify
the phase with the narrowest lines. The phase 0.09 observation
was determined to have the sharpest lines, thereby representing the spectrum
that was least affected by magnetic broadening.
The program SYNTHE and atomic line data of Kurucz were then
used to generate synthetic spectra. The ATLAS9 model having
the atmospheric parameters T$_{\rm eff}$ = 11500 K, log $g$ = 4.0,
microturbulent velocity 2.0 km s$^{-1}$, and a 10 times solar metallicity 
was chosen.  By comparison of the observation with synthetic spectra it was
determined that $v\sin i$ = 14 km s$^{-1}$. This value is
smaller than previously published values: 29 (Hoffleit \& Jaschek 1982), 
24 (Wolff \& Preston 1978), 18 (Abt et al. 1972), and 17 km s$^{-1}$ 
(Khokhlova \& Pavlova 1984). 

\subsection{Elemental abundances}

The published literature features a wide range of values for the elemental
abundances of $\alpha^2$ CVn.  Perhaps the most worrisome are those for the 
iron-group elements, for which estimates range from solar values to 100 times
solar. Starting the abundance analysis at UV wavelengths is
difficult due to the line blending and the uncertainty of continuum placement.
Thus, the analysis was begun at red wavelengths, where continuum placement
was often a trivial task and line blending is minimal. 
The same atmospheric model parameters as above were used with opacities
of 1, 10, and 100 times the solar opacity.  The best fits
to the Fe\,{\sc ii} lines are for an  abundance enhancement of [Fe/H] = +0.7 at
phase 0.09 and +0.5 at phase 0.91. Other iron-group elements show similar
enhancements. The rare-earth elements (REE) dominate the bluer optical 
wavelengths,
and their enhancements were found to be more extreme, typically between
3.5 and 4.5 orders of magnitude greater than the solar values. However,
it must be cautioned that these abundance enhancements were derived using 
a line list that did not account for hfs, IS, or magnetic broadening. 
The inclusion of these effects is likely to change the abundance enhancements
determined for the REE.

\section{The UV continuum: Where is it ?}

Theoretically, if the
line list is complete and the stellar parameters are known, then synthetic 
spectra can be used to assign the continuum level. However, the absence of a
myriad of REE third spectra lines from the linelists at UV wavelengths renders
this problem more difficult than continuum placement at optical wavelengths.
This problem is illustrated in Figure 1, which compares an HST/GHRS observation
of $\alpha^2$ CVn with a synthetic spectrum 
that was calculated using the best fit $v\sin i$ and elemental abundances
from the optical region analysis. The two spectra have been arbitrarily 
normalized to 
the peak in the observed spectrum. The absence of this peak, due to either 
a slightly different grating setting or the use of lower spectral resolution
would likely lead to choosing a continuum level that is too low. Those lines
in the figure that are calculated as too strong relative to the observation 
most likely reflect either poor oscillator strengths or a residual error in 
the continuum placement. Careful study of these spectra also reveals missing
opacity due to either incorrect model elemental abundances or missing 
atomic lines in the calculation.

%%%%%%%%%%%%%%%%%%%%%%%%%%%%%%%%%%%%%%%%%%%%%%%%%%%%%%%%%%%%%%%%%%%%%%%%%%%%%%
\begin{figure}[t]
\centerline{
\psfig{figure=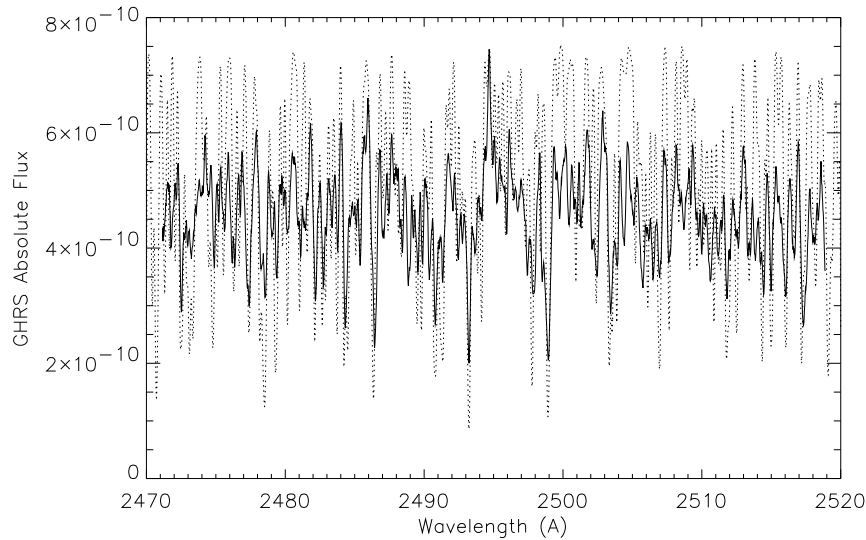,angle=90,width=11.5cm}}                                        
\caption{$\alpha^2$ CVn GHRS observation (solid) and synthetic spectrum
(dashed).}    
\end{figure}                                                                
%%%%%%%%%%%%%%%%%%%%%%%%%%%%%%%%%%%%%%%%%%%%%%%%%%%%%%%%%%%%%%%%%%%%%%%%%%%%%%

%%%%%%%%%%%%%%%%%%%%%%%%%%%%%%%%%%%%%%%%%%%%%%%%%%%%%%%%%%%%%%%%%%%%%%%%%%%%%%
%                       A C K N O W L E D G E M E N T S                      %
% Next lines show you how to write acknowledgements                          %
%%%%%%%%%%%%%%%%%%%%%%%%%%%%%%%%%%%%%%%%%%%%%%%%%%%%%%%%%%%%%%%%%%%%%%%%%%%%%%
% You must leave a blank line before the \acknowledgements command!

\acknowledgements
% Do not leave a blank line here! <---------------------->
The author gratefully acknowledges a grant from the Royal Swedish
Academy of Sciences.

%%%%%%%%%%%%%%%%%%%%%%%%%%%%%%%%%%%%%%%%%%%%%%%%%%%%%%%%%%%%%%%%%%%%%%%%%%%%%%
%                       R E F E R E N C E S                                  %
% References should start with the \begin{thebibliography}{} command, leaving%
% the last curly brackets empty.                                             %
%%%%%%%%%%%%%%%%%%%%%%%%%%%%%%%%%%%%%%%%%%%%%%%%%%%%%%%%%%%%%%%%%%%%%%%%%%%%%%

%%%%%%%%%%%%%%%%%%%%%%%%%%%%%%%%%%%%%%%%%%%%%%%%%%%%%%%%%%%%%%%%%%%%%%%%%%%%%%
%                       H A P P Y E N D                                      %
% Your LaTeX source text must be ended by the line:                          %
%%%%%%%%%%%%%%%%%%%%%%%%%%%%%%%%%%%%%%%%%%%%%%%%%%%%%%%%%%%%%%%%%%%%%%%%%%%%%%
\end{document}